\documentclass[12pt]{article}

\def\cite#1{[\ref{#1}]}
\def\citd#1#2{[\ref{#1},\ref{#2}]}

\def\ae{\alpha_{\mbox{\scriptsize eff}}}
\def\half{\mbox{\small $\frac{1}{2}$}}
\def\EEC{{\rm EEC}}
\def\MSbar{\overline{\mbox{\scriptsize MS}}}

\makeatletter
\let\chapter\hid@chapter
\makeatother
\begin{document}

\title{QCD, Theoretical issues.\\[0.4 cm]
{\normalsize plenary talk at the HEP EPS Conference,\\ Jerusalem, August 1997}}
\author{Yu. L.\, Dokshitzer\\[0.2 cm]
INFN, sezione di Milano, Italy, 
 and\\
  St. Petersburg Nuclear Physics Institute, Russia}

\date{}
\maketitle

\begin{abstract}
Today's QCD problems, prospects and achievements are reviewed.
\end{abstract}

\def\bit{\begin{itemize}}
\def\eit{\end{itemize}}
\def\nt#1{({\color{red}#1})}
\pagestyle{empty}
\def\err#1#2{\scriptsize\begin{array}{l} +0.#1 \\-0.#2\end{array}}
\def\matt#1{{\mbox{\scriptsize #1}}}
\def\refup#1{$^{[\ref{#1}]}$}
\def\fm{{\rm fm}}
\def\ben{\begin{enumerate}}
\def\een{\end{enumerate}}
\def\bit{\begin{itemize}}
\def\eit{\end{itemize}}

\def\but{\quad{\em but\quad }}

\section{Introduction: QCD in 13 puzzles}

Exploring the gap between small- and large-distance-dominated
phenomena remains the challenge for quantum chromodynamics.
QCD is the strangest construction in the history of modern physics.
Practically nobody doubts today that it {\bf is} the microscopic
theory of strong interactions; at the same time we are practically as
vague in addressing the basic questions of this Field Theory as we
were 25 years ago.

\begin{center}
{\large\bf Puzzle of} 
\end{center}

\noindent
{\bf Objects:}\quad quarks/gluons are the Truth \but hadrons are the Reality. 

\noindent
{\bf Rules:}\quad   sacred \but\ absent. 
The QCD Lagrangian is a beautiful construction.
At the microscopic level, {\em bal tosif, bal tigra}~\cite{Torah}. 
No initiative is allowed, as in a FSU ``administrative economy''.
When it comes to the macroscopic hadron world, on the contrary, no
rules seem to exist. As in a market economy, what matters is not the
quality of the product, but whether you manage to fool your customer
into buying it.

\noindent
{\bf Responsibility:}\quad gluons are essential \but quarks dominate.
The non-Abelian gluon self-interaction is crucial for asymptotic
freedom and thus for causing an infrared instability. However, the
hadron world as we know it remains persistently quark-driven.

\noindent
{\bf Scales:} finite interaction radius \but $m_{gluon}\equiv 0$.
No field-theoretical mechanism is known that would protect a theory
possessing a strictly massless gluon field 
from developing long-range forces.  

\noindent
{\bf Binding, Relativity, Multiplicity:}\quad  
we deal with strongly interacting practically massless
particles, hence one could expect many-body ($N\!\gg\!1$) relativistic
($1\!-\!v\!\ll\!1$) dynamics,
 \but\ we see additive quarks, surprising successes
of the non-relativistic valence quark model. 

\noindent
{\bf Goals \&\ Means:} \quad We aim at describing a colourless 
world in terms of propagation and interaction of coloured quarks and gluons. 
Strictly speaking, we do not know how to do it consistently, that is how to 
properly define gluon degrees of freedom, even within perturbation theory. 
The problem of ``Gribov copies'' is there and remains unsolved.

\noindent
{\bf Perturbation:}\quad $\alpha_s(Q^2)$ is small 
\but\ often all orders are essential.
Depending on the problem under consideration, the true perturbative
expansion parameter is often enhanced by power(s) of $\log Q^2$.  

\noindent
{\bf Evolution:}\quad Quantum Field Theory \but\ classical probabilities.

\noindent
{\bf Freedom:}\quad quarks imprisoned 
          \but\ fly away for about $10^3$ fm these days.

\noindent
{\bf Confinement:}\quad Inevitable \but\ elusive. 
Certain global characteristics of multi-hadron production in hard
processes, such as inclusive energy spectra of light hadrons (pions,
all charged) in jets, multiplicity hadron flows in the inter-jet regions, 
show no sign of hadronisation effects.

\noindent
{\bf Coupling:}\quad ``Strong interactions'' \but 
$\left\langle \frac{\alpha_s}{\pi}\right\rangle_{\mbox{\scriptsize
infrared.}} \sim 0.2\>.$

``Confinement? What's the problem? 
Hasn't it been solved long ago by W. [\,the name may vary\,]?''
--- would be a mean theoretician's response. 
As a reflection of this attitude, the authors 
of experimental papers dealing with hadroproduction increasingly 
often tend to equate QCD with Monte Carlo hadron generators. 
Not only does time cure: it also tends to put out of focus, for the sake
of psychological protection, difficult problems that have resisted
head-on theoretical attacks for far too long. 
We may appreciate the former property of time; we'd better be alarmed by
the latter. 

Vladimir Gribov, a brilliant physicist and a strong character, had 
enough curiosity, motivation and stamina to pursue a non-stop 20-year 
study of QCD confinement, till that night of August 13,
1997, when he passed away.

The founder and the leader of the renouned ``Leningrad school of theoretical
physics'', Gribov belongs to the generation of physicists, 
now almost extinct, 
who did not take the Quantum Field Theory (QFT) for granted. 
Thus he was able to raise heretic questions, putting under
scrutiny the very basics of the QFT as we (think we) know it. 
Let me list for you some of them:
\bit
\item
Are the three approaches  --- secondary quantisation, Feynman
diagrams, functional integrals --- really equivalent?

\item
Does classical topology play any r\^ole in QFT?

\item
Can Euclidean rotation be justified and thus a statistical system 
substituted for QFT in the case of infrared unstable dynamics?

\item
How should an ultraviolet renormalization programme be carried out in
a theory where the physical states do not resemble the fundamental
fields the Lagrangian is made of?
  
\item
Whether the Dirac picture of the vacuum, with negative-energy levels
being occupied and the positive-energy ones being kept empty, 
applies to quarks?

\item 
How to bind together massless particles?

\item
Shouldn't the electro-weak symmetry breaking scale and the QCD scale be
related with one another by a pion which is an electro-weak point-like
Goldstone and, at the same time, 
a quark-antiquark bound state from the QCD point of view?

\eit
In general, the confinement problem is a problem of understanding 
and describing the physical states of a Field Theory with (unbroken) 
non-Abelian gauge symmetry. There can be many confinements, that is
many solutions to the non-Abelian instability. 
The core observation made by V.N.~Gribov was that {\em the}\/
confinement, that is confinement in the world we live and experiment
in, is largely determined by the fact that   
practically massless quarks are present in the theory. 

Gribov's ideas remain to be discovered, understood and developed.

\section{Small $x$ and Confinement (Heron vs. Pomeron)}

{\bf A message from DIS 97, Chicago:}

A rare thing is more damaging to empirical science than an improper name.
In perturbative QCD we may discuss the BFKL {\em approximation}\/, 
the BFKL {\em equation}\/, 
BFKL {\em dynamics}\/ but should 
restrain from talking about a ``BFKL or Hard Pomeron''. 
There are two reasons for that. First of all, the Pomeron 
(Gell-Mann's name for the Gribov vacuum pole or vacuum singularity)
is a reserved word... 
More importantly, the very term ``Hard Pomeron'' is nonsensical. 
\begin{itemize}
\item {\bf Pomeron:} 
  a leading high-energy contribution to elastic hadron scattering
  amplitudes driven by the leading singularity in the $t$-channel 
  (complex) angular momentum $\omega$. 
\item{\bf Hard:}  
  determined by small distances, and therefore perturbatively controllable. 
\end{itemize}
These two word simply don't merge, since the position and the nature of the
leading singularity in $\omega$ is entirely off the books of perturbative QCD.

\subsection{BFKL and confinement}

\begin{flushright}
\begin{minipage}{6.5 cm}
``I am but mad north-north-west: 
 when the wind is southerly I 
 know a hawk from a handsaw.''

\hfill Hamlet, Prince of Denmark
\end{minipage}
\end{flushright}
 
The fact that the behaviour of DIS structure functions at fixed $Q^2$ and
$x\to0$ is an entirely {\em non}\/-perturbative phenomenon, was
recently demonstrated, beyond any reasonable doubt, by 
Camici \&\ Ciafaloni and by A.~Mueller.

In Mueller's paper~\cite{Mueller} an estimate is given of the range of $x$
beyond which the perturbative treatment (operator product expansion,
OPE) breaks down due to diffusion into the small-transverse-momentum 
(strong interaction) region. 
Numerically,
$$  
    Y= \ln {x}^{-1} \>\ll\> \left({2\alpha_s(Q_0)}\right)^{-3} ,
$$
with $Q_0$ the {\em minimal}\/ of the two scales, to which   
high energy $s\propto 1/x$ is applied. 
This leaves no hope for 
{\em predicting}\/ the $x$-behaviour of the
initial parton distributions at virtualities as low as 1--2 GeV,
and, therefore, the $x$-behaviour of the DIS structure functions. 

Camici and Ciafaloni have shown in \cite{CCnpt}, in a very nice
physically transparent way, that the position and the nature of the
leading complex-angular-momentum singularity (Pomeron) depend
crucially on the behaviour of the QCD coupling $\alpha_s$ in the infrared
region. Read: have non-perturbative origin. 

This does not mean that the BFKL analysis and results are nice but irrelevant.
We have to come to terms with the fact that the BFKL-predicted
increase of cross sections with energy is of little
relevance for DIS structure functions. 
However, in special circumstances it remains a perfectly sound 
and unquestionable QCD prediction which should be seen experimentally. 

Instead of the``BFKL Pomeron'' we may say the BFKL {\bf HER}on, a temporary  
{\bf H}igh-{\bf E}nergy {\bf R}egime
of increasing {\em small}\/ interaction cross sections of 
{\em small}\/ hadronic objects.

\subsection{Next-to-leading BFKL dynamics}
The next-to-leading analysis of the BFKL problem is about to be
completed. The necessary ingredients have been
computed, in recent years, in a series of highly technical works by 
Camici, Ciafaloni, Fadin, Kotsky, Lipatov, Quartarolo. 
What remains to be done is 
to deduce how the BFKL equation is modified beyond leading
logarithms, whether the Regge-Gribov factorisation is respected, 
how the coupling runs in the BFKL kernel, how much down the energy-growth
exponent (the ``BFKL intercept'') goes. 
These questions are not easy to answer, even though all the
next-to-leading corrections, to this and that, are known. 
To put it simple, a severe ``ideological'' problem remains of how 
to {\em formulate}\/ the improved answer. 
For example, the very notion of the BFKL exponent $\lambda$ is elusive: 
    how to quantify 
   ``{\em asymptotic behaviour}''  of a {\em non-asymptotic}\/ amplitude? 
(The true asymptotia, as we know, is the soft = non-perturbative Pomeron.) 

Let us present the behaviour of the forward amplitude (total cross
section) for the scattering of two objects with small sizes 
$p_t^{-1}\ll 1\fm$  at invariant energy $\hat{s}$  as
$$ 
 \mbox{Heron}(\hat{s},p_t^2;0) \propto \left(\frac{\hat{s}}{p_t^2}\right)^
 {\lambda(\alpha_s(p_t^2))}.
$$
(The third argument of $H$ is the momentum transfer along the
BFKL-ladder, $t=0$ for the forward scattering amplitude.)

A {\em preliminary}\/ result for the Heron ``intercept'' 1997 is~\cite{CCint}
$$
 \lambda^{1997}(\alpha_s)= \lambda^{1976}(\alpha_s)\cdot
 \left(1-3.4N_c \frac{\alpha_s}{\pi} - 0.15 n_f\frac{\alpha_s}{\pi}
  \>+\> {\cal{O}}(\alpha_s^2)\right). 
$$
Numerically,
$$
 \lambda = 0.5 \>\>\to \>\> \lambda=0.2 \qquad \mbox{for}\>\>\alpha_s=0.15  
\>\>\>{(?)}
$$
If we close our eyes on the inapplicability of the BFKL Heron to DIS structure
functions 
and boldly translate the next-to-leading BFKL correction into the DIS
anomalous dimension,  weird results may follow.
As was demonstrated by Bl\"umlein \&\ Vogt\footnote{The subject field
  of the Johannes Bl\"umlein's email message read 
  ``suicide of small-x physics''}, the gluon
splitting yields negative probability, structure functions 
(a gluon-driven $F_L$ in the first place) start decreasing at small $x$,
and the like~\cite{BV}. 
Not that the BFKL-corrected anomalous dimension adopted by B\&V
was free of criticism~\cite{Marcello_pc}. 
What seems to me a more important shortfall is asking a wrong question. 
Let DIS SF rest in peace!

The crisis is over. Back to work.

\subsection{Two large-and-equal scales} 

In \underline{$ep$-scattering} we look at forward jet production 
(Mueller-Navelet jets) with jet transverse momentum $p_t^2\simeq
Q^2$. This is a two-large-scale problem, and the corresponding cross
section should increase with the boson--jet pair energy as 
$$
 d\sigma^{\mbox{\scriptsize MN}} \propto
 \mbox{Heron}(\hat{s},Q^2;0),\qquad
\hat{s}=Q^2x_{jet}/x_{Bjorken}\gg Q^2\,\>.
$$

\underline{In $pp$-scattering} we may discuss at least two options.
The first is {\em inclusive}\/ production of a high-invariant-mass 
pair of jets with large (and comparable) transverse momenta,
$s\gg\hat{s}=(p_1+p_2)^2\gg p_t^2\>$. 
In this case we expect
$$
 d\sigma^{\mbox{\scriptsize incl.}} \propto
 \mbox{Heron}(\hat{s},p_t^2;0)\,.
$$
If we impose a condition of having a {\em rapidity gap}\/ between the
  jets, then for rapidities large enough so as to eliminate the
  Sudakov-suppressed one-gluon exchange, we should have
$$
     d\sigma^{\mbox{\scriptsize gap}} \propto 
      \left(\mbox{Heron}(\hat{s},p_t^2; -p_t^2)\right)^2.
$$
(Recall, the last argument of the Heron amplitude is the momentum
transfer; here $t\approx-p_t^2$.)
The latter ``double-diffractive'' process is a unitarity-shadow of the
former (total, inclusive) one.

The Heron may also be searched for in \underline{$\gamma\gamma$ collisions} 
in conditions similar to $pp$, as well as in double-hard
double-diffractive processes like the ``dream process'' 
$\gamma\gamma\to J/\psi+J/\psi$,
$$
     d\sigma_{\gamma\gamma\to J/\psi J/\psi}
      ^{\mbox{\scriptsize frwrd.,excl.}} \propto 
      \left(\mbox{Heron}(\hat{s},m_c^2; 0)\right)^2\>.
$$

Two comments are due concerning experimental BFKL searches, one
optimistic, another rather pessimistic.

\subsubsection{Mueller-Navelet experiment and ``BFKL at hadron level''.}
ZEUS exercises some caution before claiming that the
BFKL-predicted increase of MN-jet cross section is observed 
experimentally~\cite{ZEUS_MN}. 
This caution is well grounded but may be slightly exaggerated. 
As I understand, it stems from the observation that the standard
cascade-type generators (Lepto, Herwig) exhibit an alarmingly large
mismatch between the parton- and hadron-level cross sections. 

However, this does not necessarily imply that the lacking hadron-level BFKL
prediction is really necessary. 
The very idea of jet finding algorithms was to keep the correspondence
between the parton and hadron ensembles. 
The fact that the MC generators fail to preserve this 
correspondence may be simply due to their inability to properly 
estimate the parton level cross section:
they can produce the MN-events only as improbable fluctuations.

Indeed, a starting point for BFKL in the MN-experiment is the 
$\alpha_s^2$ QCD matrix element.  
For small values of Bjorken $x$, 
production of a forward jet with $p_\perp^2\sim Q^2$
is accompanied by two more quark jets with transverse momenta of the
same order, coming from the Boson-Gluon-Fusion box. 
None of the standard MC generators based on the logarithmic DGLAP evolution 
picture ever pretended to embody such high-order configurations with 3
partons at the same hardness scale (corresponding to the two-loop
correction to the coefficient function).

Therefore, MC models are likely to be responsible for the mismatch between 
parton- and hadron-level results, not the physics, which is the BFKL
physics in this case.

\subsubsection{Azimuthal de-correlation as a sign 
of the BFKL dynamics.} 
Accompanying gluon radiation in the production of 
two jets with large $p_t$ has a double-logarithmic nature. 
It consists of gluons with transverse
momenta distributed logarithmically from small $k_{ti}^2\ll p_t^2$ 
up to $k_{ti}^2\sim p_t^2$ at a given rapidity (energy log) 
and uniformly in rapidity (angular log). In the inclusive two-jet
cross section the double-logs disappear (real and virtual gluons with 
$k_{ti}^2\ll p_t^2$ cancel). 
The result can be expressed in terms of (an imaginary part of) 
the single-logarithmic forward amplitude known as BFKL. 

Studying the $\hat{s}$ dependence of the two-jet cross section,
$\hat{s}=x_1x_2s$ (keeping the $x_1$, $x_2$ fixed as to factor out the
initial parton distributions) is a straight road to verifying the BFKL
dynamics.
However, the cancellation of small-$k_\perp$ gluons mentioned above
is not present in less inclusive quantities 
such as the distribution in the total transverse momentum of the two
triggered jets, or in the azimuthal angle between the jets, 
in particular in the back-to-back region, $\Delta\phi=\pi-\phi_{12}\ll 1$.  
Sudakov form factor suppression broadens the 
$\Delta\phi$--distribution.
Moreover, this broadening (azimuthal de-correlation) 
increases with $\hat{s}$ roughly as
$$
  \frac{d\sigma}{d\phi} \propto
(\Delta\phi)^{-1+   N_c\frac{ \alpha_s}{\pi}
\ln\frac{\hat{s}}{p_\perp^2}}\,.
$$
The $\hat{s}$--dependent part of the de-correlation 
is due to the dynamical suppression of soft gluon emission at ``large
angles'', that is within the broad rapidity interval between the jets. 
Since in two-jet production at large $\hat{s}$ one-gluon exchange dominates, 
this coherent radiation is determined by the colour charge of
the $t$-channel exchange, i.e. by that of the gluon. Hence, the $N_c$
factor in the radiation intensity. It is related with QCD Reggeization
of the {\em gluon}\/ and has little to do with the vacuum-exchange
dynamics, that is with the BFKL phenomenon. 

There is a hope to see the BFKL-motivated energy dependence of the first
{\em moments}\/ of the $\phi_{12}$--distribution, in which the
back-to-back region is suppressed~\citd{Delduca}{D0azim}. 
This is not impossible. However, the necessity of hunting down single-log
effects in the presence of double-logs in the differential
distribution makes one feel suspicious about this option of
visualising the BFKL dynamics.

\subsection{0- and 1-scale diffractive (quasi-elastic) processes} 

The following HERA chart illustrates the transition from 0-- to
1--hard-scale processes:
$$
\begin{array}{|c|c|} \hline & \\
\sigma_\matt{tot}(\gamma p) &  \mbox{Pomeron} \\[2mm]
\sigma_\matt{D}(\gamma p) &  [\mbox{Pomeron}]^2 \\[2mm]
\hline 
\sigma_\matt{tot}(\gamma^*p) &   G(x,Q^2) \\[2mm]
\sigma_\matt{D}(\gamma^*p) &   G^2(x,Q^2) \\[2mm]
\sigma(\gamma^{(*)} p\to J/\psi +p^*) 
& [G(x,m_c^2)]^2\to [G(x,Q^2)]^2 \\[2mm]
\sigma(\gamma^{(*)} p\to \mbox{2jets}(k_t)+ p^*) 
& [G(x,k_t^2)]^2\to [G(x,Q^2)]^2 \\ [2mm] 
\hline
\end{array}
$$
In the second block the r\^ole of a hard scale may be played 
by the initial photon virtuality $Q^2$, a heavy quark mass, or large 
{\em transverse momenta}\/ inside a diffractively produced system. 
(Notice that a large {\em invariant mass}\/ of the latter does not
qualify.)

In the two last cases $\gamma p$ and $\gamma^* p$ (DIS) stand on an equal
footing. With increasing photon virtuality, $Q^2$ may take over
from the quark mass or the jet transverse momentum as the hardness argument
determining the scale of the gluon distribution. 

Here there is no room for the Heron amplitude which applies to
the scattering of {\bf two} small objects:

\noindent
The energy dependence of 
total (inclusive) cross sections are, for the 

scattering of 
$\left\{
\mbox{\begin{minipage}{5in}
Large on Large: \quad {\bf Pomeron},\\
Small on Large: \quad {\bf Gluon} {(in the proton)}, \\
Small on Small: \quad {\bf Heron}.
\end{minipage}
}\right.$

\vspace {3 mm}
\noindent
The energy dependence of 
diffractive cross sections are, for the

scattering of 
$\left\{
\mbox{\begin{minipage}{5in}
Large on Large: \quad ({\bf Pomeron})$^2$,\\
Small on Large: \quad ({\bf Gluon})$^2$, \\
Small on Small: \quad ({\bf Heron})$^2$.
\end{minipage}
}\right.$

\vspace {0.5 cm}

\noindent
There is an interesting option to ensure two small scales, and thus to 
approach BFKL dynamics, namely by studying diffractive processes, 
e.g. $\gamma p \to \rho +p^*$ at sufficiently large momentum transfer $t$.
Forshaw and Ryskin have recently verified that a finite momentum
transfer $|t|$ suppresses large-distance-directed 
diffusion and thus keeps the BFKL gluon system under perturbative
control. 
For example, one can expect the two-gluon exchange to turn into 
a fully-fleshed Heron
in the $s/{t}\gg1$ limit of (double-) diffraction processes
with $|t|>$~few~GeV$^2$,
$$
\begin{array}{|c|c|}
\hline & \\
 \gamma p \to \gamma^{(**)}+p^{(*)} & 2g\to [\mbox{Heron}(\hat{s},0;t)]^2
 \\ & \\
\hline
\end{array} 
$$
where $\gamma^{(**)}= \gamma,\> \rho,\> J/\psi,  Z^0$, whatever.

\section{Coherence and the Nucleus as Colorometer}
Our field has emerged as a result of the digression: 
natural philosophy $\to$ physics $\to$ quantum physics $\to$ 
elementary particle physics. 
The older generation participated in the next step,  
elementary particle physics $\to$ high energy physics. 
In the past 20 years we have witnessed the final split \\
high energy physics 
$\>\>\to\>\>$ $\left\{\> \mbox{\begin{minipage}{1.0in} soft physics \\
                        hard physics \end{minipage}} \right. .$ \\
It is about time to restore the integrity of the subject.
{\bf\em Coherence} 
is the key-word for re-integrating {\em soft}\/ and {\em hard}\/ 
physics into {\em high energy physics}.

You can smell new {\bf quasi-classics} in the air
when you hear discussions of 
high gluon densities (hot spots), disoriented chiral condensates, 
percolating strings, quark-gluon plasmas, the effective colour field 
of a nucleus, etc.
What remains to be done is 
to condense this smell into a marketable fragrance. 

High-energy hadron-scattering phenomena may be more {\em classical}\/
than we use to think they are.
To illustrate the point let us look into Scattering phenomenon and
High Energies.  

\subsection{Scattering}
V.N.~Gribov produced an argument in favour of
Scattering in the Quantum Field Theory framework being an inch closer 
to Classical Scattering than, paradoxically,  that in Quantum
Mechanics.
He addressed the question 
``Why does the total cross section rise in the first place?''

Total cross sections of classical potential scattering are typically infinite.
For example, the tail of the Yukava potential, 
$V(r) = \mu \exp{(-\mu r)}/r$,
produces small-angle scattering which makes $\sigma_{tot}=\infty$.

What keeps $\sigma_{tot}$ finite in quantum mechanics, as we know, is 
the impossibility to separate a go-straight wave from a diffracted one 
in the case of too small a scattering angle, $p_\perp=p\theta<1/\rho$. 

QFT, however, gives us such an option by introducing 
{\em inelastic diffraction}. The final state being different from the
initial one, the QM argument breaks down, and large impact parameters 
start contributing to the total cross section. The latter increases 
with energy due to the decrease of $t_\matt{min}$ with $s$.

\subsection{High Energies}
The hadron as a QFT object is a coherent sum of various configurations.
The quantum portrait of a projectile, its field fluctuation, 
stays frozen in the course of interaction at high energies, so that
each fluctuation scatters independently~\cite{FP}.
The total interaction cross section, for example, emerges
as a classical (incoherent) sum of cross sections of different configurations
inside the projectile hadron $h$, each of which interacts with its private
$\sigma$. 
Introducing the corresponding distribution~\cite{GW}, 
we may write
$$
 \sigma^h_\matt{tot} \>=\>\left\langle\sigma\right\rangle_h 
 \equiv \int d\sigma\> \sigma\cdot p_h(\sigma)\>.
$$
The issue of coherence comes onto the stage when we turn from the total
cross sections to more subtle phenomena, {\bf diffraction} being
rightfully the first among many.

Let us repeat,  
an incident proton is a coherent sum of different field fluctuations. 
Talking hadrons, they are $p\to \pi^+ n\to p$, $p\to K^+\Lambda\to p$,
etc. Switching to the quark language, we may talk about three valence
quarks at various relative distances, with a correspondingly suppressed
or enhanced gluon field between them, that is to picture the proton as
being virtually squeezed or swollen. 
As was noticed by Feinberg and Pomeranchuk, 
if all the configurations inside the projectile 
interacted {\em identically}\/ with the target, such interaction would not
induce {\em inelastic diffraction}. 
Indeed, in this case the interaction Hamiltonian  
preserves the coherence of the initial state which, therefore, 
would scatter elastically but would not break apart producing 
small-mass diffractive states. 
In other words, inelastic diffraction is sensitive 
to the $\sigma$-{\em distribution}.  
The dispersion of the latter can be directly measured in diffraction on
nuclei~\cite{MP}
$$
\left. \frac{\sigma(h A\to h^{*}A)}{\sigma(h A\to hA)}\right|_{t=0}
= \frac{\left\langle\sigma^2\right\rangle_h}
{\left\langle\sigma\right\rangle_h^2} -1\>.
$$
In general, high energy scattering off nuclear targets provides
indispensable tools for studying the internal structure of hadrons. 

\subsubsection{Proton-penetrator.}
The $A$-dependence of inelastic diffraction 
gives a classical example of the use of the nucleus as a {\em colorometer}.
If an incident hadron ($p$, $\pi$) {\em always}\/ interacted strongly
with the target in central collisions, inelastic diffraction would be 
peripheral and therefore its cross section would grow as $A^{1/3}$
(total absorption being an example of an ``identical interaction'').
Experimentally (FNAL, ISR), $\sigma_D\sim A^{0.8}$.
This shows that there are {\em small}\/ configurations inside the
proton that penetrate the nucleus.  
It is not easy to visualise such a configuration (``penetrator'')
using the hadronic fluctuation picture. 
However, it is readily understood
in the quark language: tightly placed valence quarks with abnormally
small gluon field between them, due to quasi-local compensation of
the colour charge. 

The interaction of such fluctuations with the target stays under
the jurisdiction of pQCD. Theoretically, there is a nice tool for describing
the interaction of a simpler projectile, that is 
a tight colourless $q\bar{q}$ pair: 
$$
 \sigma(q\bar{q},A) = \frac{\pi^2}{3}\,b^2\, \alpha_s(b^{-2})\>
xG_A(x,b^{-2})\>.
$$
It relates the interaction cross section of a quark and an antiquark 
at small relative transverse distance $b$ (inside a real/virtual
photon or a meson)
with the gluon distribution in the target~\cite{FRS}. 
Its application to the $A$- and $s$-dependence of shadowing, to colour 
transparency phenomena, and many other intriguing subjects marked the
beginning of QCD ``Geometric colour optics'', the name of the game
invented by Frankfurt and Strikman.

\subsubsection{Proton-perpetrator.}
Thus, diffraction and transparency phenomena can be employed to study 
squeezed, penetrating hadrons. Their opposite members --- swollen
hadrons --- can also be visualised in scattering on nuclei.

Contrary to the penetrator, the proton-perpetrator should willingly interact
with the target. In such configurations  ``strings'' are pulled, 
colour (or, pion) fields are stronger than typical, the vacuum is virtually   
broken. So, one can expect an enhanced yield of antiquarks pre-prepared to 
be released when coherence of the projectile is destroyed. 
Perpetrators can take responsibility for a bunch of puzzling phenomena 
observed in $pA$ and ion-ion ($AB$) collisions, 
such as baryon stopping, the increase of the $\Lambda/p$ ratio (Kharzeev)
and an amazingly large relative yield of {\em antibaryons}.

To enhance the r\^ole of non-typical configurations one should 
look at the {\em tails}\/ of various distributions. 
In particular, by looking into events with {\em very large}\/ $E_t$ in ion-ion 
collisions one may see unusual percolated nuclei consisting of 
merged nucleons, instead of the much too familiar nucleon gas.
The quark-gluon plasma state is usually thought to be formed in the collision, 
which provides a melting pot for individual nucleons. A large $E_t$-yield 
is considered to be a sign of the phase transition into such a state, 
in the course of the collision.  
Alternatively, very large $E_t$ may be looked upon as a {\em precondition}\/
for the collision, rather than the result of it: to observe a larger than 
typical transverse energy yield, we catch the colliding nuclei in rare, 
confinement-perpetrating, virtually melted configurations a'la the desired 
plasma state.  

In such configurations the yield of Drell-Yan lepton pairs should be higher 
(extra pions, or antiquarks, around); 
the reduced yield of $J/\psi$ (heavy traffic) is to be expected.

\subsection{Praising $J/\psi$ ---  a boarder-guard} 
Heavy onia remain a pseudo-perturbative tool for probing strong interaction 
dynamics, and a source of surprises at the same time. 
One of the striking puzzles is offered by the comparison of 
{\em medium-dependent}\/ transverse momentum broadening of Drell-Yan pairs, 
$J/\psi$ and $\Upsilon$ observed in $pA$ collisions: 
$\ell^+\ell^-:J/\psi:\Upsilon\simeq 1:2:8$.
Future explanation of this puzzle should shade light on the production 
mechanism of vector $Q\bar{Q}$ onia. 

The cross section for $J/\psi$ production in $pp$ collisions is  
a long-standing problem: the pQCD treatment based on the gluon-gluon fusion 
into $J/\psi$ plus the final-state gluon, $gg\to J/\psi\!+\!g$, is way off.

To avoid an extra $\alpha_s$-cost the so-called ``octet mechanism'' was 
suggested based on producing a colour-{\em octet}\/ $c\bar{c}$
which then gets rid of colour by radiating off ``a soft, non-PT gluon'', for 
free. This quite popular picture, however, misses one essential point: 
According to the renouned Low-Barnet-Kroll theorem, soft radiation does 
not change the state of the radiating system. 
In other words, soft radiation does not carry away quantum numbers. 
Colour is no exception: paradoxical though it may sound, 
soft gluons cannot blanch the octet $c\bar{c}$ into a white 
$J/\psi$. 

Physically, soft gluons are radiated {\em coherently}\/ by the colour
charges involved, and enter the stage, formally speaking, long after 
the $J/\psi$ formation time. Technically, a soft gluon, that is one with 
the characteristic logarithmic energy spectrum  $d\omega/\omega$, sits, 
colour-wise,  in an {\em antisymmetric}\/ $f_{abc}$ configuration with its 
two partners. However, to form a $C$-odd $J/\psi$ state one needs a   
{\em symmetric}\/  $d_{abc}$ colour coupling instead. This is the reason why 
in the decay channel, $J/\psi\to ggg$, none of the gluons is allowed to be 
soft: in the small energy domain one has the $\omega d\omega$ distribution 
rather than the classical $d\omega/\omega$.  

There is no doubt that $J/\psi$ is not an entirely small-distance object,
not a pure non-relativistic Coulomb $c\bar{c}$ system.
Therefore the idea of introducing into the game new non-perturbative 
matrix elements, in other words, large-distance configurations in the $J/\psi$ 
field wave function (Braaten et al) is well grounded. What makes me feel 
uneasy, however, is an accompanying (hopefully, unnecessary) chant 
of the free-of-charge soft-gluon blanching of the system, which sounds too
much like a free lunch.

\section{QCD checks 1997}
Let me start by citing George Sterman 
at the DIS conference in Chicago last spring:
\begin{quote}
``The study of quantum chromodynamics and the investigation
of hadronic scattering are the most challenging problems
in quantum field theory that are currently accessible in the laboratory.''
\end{quote}

\subsection{QCD in danger}

A decade after the JADE collaboration, 
which pioneered the $e^+e^-$ jet studies at PETRA, 
was dissolved, impressive new JADE results have appeared.  
Reanalysis of the old data was necessitated by the new jet measure
(broadening), new jet finder (Durham) and new wisdom (confinement
$1/Q$ effects in jet shape observables, see below).
  
Have a look at few citations from 
``A study of Event Shapes and Determination of $\alpha_s$
using data of $e^+e^-$ Annihilation at $\sqrt{s}$=22 to 44 GeV'', and you will
agree that the paper could have been rightfully submitted to a journal 
on Archaeology.  

\begin{quote}
%``Data recordered by the JADE experiment at the PETRA 
%$e^+e^-$ collider were used
%to measure the event shape distributions of the thrust, the heavy jet mass, 
%the wide and total \underline{jet broadening} and the 
%differential 2-jet rate in the \underline{Durham} scheme. 
%For the latter three observables, no experimental data were 
%available at these energies so far... ''

 ``The retrieval of the data eleven years after shutdown of the experiment
turned out to be \underline{cumbersome} and finally \underline{incomplete}.''

``... \underline{missing data sets} of about 250 events
 around 22 GeV and about 450  events around 44 GeV. 
In addition, the original files containing information
about the luminosity of different running periods 
\underline{could not be retrieved}...'' 
\end{quote}
Even the Acknowledgement section sounds alarming:
\begin{quote}
``We thank the DESY computer centre for copying old IBM format tapes to modern
data storage devices before the shutdown of the DESY-IBM. We also acknowledge
the effort...  
to \underline{search for} files and tapes...'' 
\end{quote}

Hadron physics is forever, because today's devotion to  
high energies is {\em temporary}. 
High energies let us watch the vacuum being excited 
and think about it for some (Lorentz-dilatated) time. 
Once the vacuum structure has been understood (the most important and
the most difficult step still to make), hadron physics will turn back to   
small and medium energies.  

Anyone willing to revisit celebrated ISR data? Forget it!  
What makes the old experiments dead as a doornail?   
\begin{enumerate}
\item the data is gone;
\item the tapes are kept somewhere, but cannot be read out
  (the recording media have changed);  
\item the data can be retrieved but nobody  
  remembers how the information 
  storage was organised: the students who knew are no longer
  around.
\end{enumerate}
To make you shiver, just imagine a similar fate awaiting the LEP-1 data!
We will never have anything comparable with the $Z$-hadron-physics-factory in 
the future. No one has enough imagination to foresee what sort of questions 
theoreticians will fancy in ten years from now. 
It will be a major catastrophe for the field if a well organised
representative set of pre-processed LEP-1 data is not available in the
future, for theoreticians to probe new ideas against the wealth of
hadronic LEP-1 information. 
A special task force should be established at CERN 
to carry out the project.

\subsection{Basic stuff}
George Sterman again:
\begin{quote}
``Perturbative methods have been found to be surprisingly, sometimes 
{\em amazingly}, flexible, when the {\em right questions}\/
are asked.'' 
\end{quote}
I took the liberty to emphasise the two ingredients crucial for the  
discussion that follows, which are: ``amazing'' and ``right questions''.

LEP and SLAC $e^+e^-$ experiments have reached a high level of
sophistication. These days they talk about identified particles 
in perfectly identified (heavy quark, light quark, gluon) jets.  
Theoretical expectations of gluon jets being softer and broader are
verified, even an elusive $C_A/C_F$ ratio is now well in place.
Extracting gluon jets from three-jet events is a tricky business which
involves choosing a proper event-geometry-dependent hardness scale 
to describe the gluon subjet (see, in particular,~\cite{DELPHI_545});  
$e^+e^-$ annihilation events with gluon jet recoiling against heavy
$Q\bar{Q}$ flying into the opposite hemisphere, though rare, provide  
a bias-free environment for studying glue~\cite{OPALglue}.

The HERA experiments are catching up, the global properties of the struck
quark jet (inclusive energy spectra and the scaling violation pattern,
KNO multiplicity fluctuations, etc.) 
converging with those of quark jets seen in $e^+e^-$~\citd{H1_251}{zeus_662}.
What makes the HERA jet studies even more exciting, 
is an ability to scan through the moderate (and small) $Q^2$ range 
in order to shed some light onto the transition between 
hard and soft phenomena.

A comparative study of the yield of different hadron species in quark
and gluon jets is underway. 
Gluon jets are reportedly richer with baryons (as was expected from the
times of hadronic $\Upsilon$ decays).
The yield of $\eta$ and $\eta'$ mesons in gluon jets is also under focus 
(L3, ALEPH). 
The numbers are still uncertain. For example, the relative excess of $\Lambda$
hyperons was reported from almost none~\cite{L3lam} up to 40\%~\cite{OPALlam}. 
There is no doubt that in the near future the situation 
will be clarified\footnote{I remember hearing during the conference 
of a 100\%\ $\Lambda$--excess, but failed to find any documented evidence.}. 

From the theoretical side, one warning is due. 
According to the present-day wisdom, the production of 
accompanying hadrons with {\em relatively small}\/ momenta  
in jets is always ``{\em gluonic}'': it is driven by multiple
radiation of soft gluons off the primary hard quark/gluon parton
which determining the nature of the jet. The gluon radiation being 
{\em universal}, so should be the relative abundance of different
hadron species in the ``sea''. 

From this point of view, the difference in the yield of hadrons 
should be there only in the leading-parton fragmentation region:
hadron fragmentation of the valence quark can differ from that of a gluon. 
So, the crucial question is: whether the differences between the quark- and
gluon-initiated jets is concentrated near the tip of the jet.
If it is not, that would be evidence for (unexpected)
gluon density effects in the dynamics of hadronisation.

\subsection{Subtleties} 
The effects of QCD coherence in hadron flows {\bf in} and 
{\bf in-between} jets are well established experimentally. 
Gluon coherence inside jets leads to the so-called ``hump-backed'' plateau in 
one-particle inclusive energy spectra.
A quantitative theoretical prediction known as the ``MLLA-LPHD prediction'' 
was derived in 1984. 
It has survived the LEP-1 scrutiny; more recently, it has been 
confirmed by a detailed CDF analysis~\cite{CDF}; 
these days it is seen also at HERA\cite{H1_251}.

This QCD prediction has two ingredients. 

\begin{itemize}
\item[MLLA] 
 stands for the ``modified leading log approximation'' of pQCD 
 and represents, in a certain sense, the resummed 
 next-to-leading-order approximation. 
 This step is necessary for deriving {\em asymptotically correct}\/ 
 predictions concerning multiple particle production in jets. 
 This means that the MLLA parton-level predictions become exact 
 in the $W^2\to\infty$ limit. 
\item[LPHD] (local parton-hadron duality) is a hypothesis rather than
  a solid QCD prediction. It was based on the idea of soft confinement,
  motivated by the analysis of the space-time picture of 
  parton multiplication, 
  and stated that observable spectra of hadrons should be
  mathematically similar to the calculated spectrum of partons
  (the bulk of which are relatively soft gluons). 
\end{itemize}
Experiment does respect LPHD~\cite{KO}. 

What makes the story really surprising is that the
perturbative QCD spectrum is mirrored by that of the pions (which constitute
90\%\ of all charged hadrons produced in jets), even at momenta below 1 GeV!
Moreover, the ratios of particle flows in the inter-jet regions in
various multi-jet configurations,
which reveal the so-called ``string'' or ``drag'' effects also respect
the parameter-free pQCD predictions based on the coherent soft gluon
radiation picture. These observables are dominated, at present
energies, by junky pions in the 100--300 MeV momentum range!

Is there any sense in applying the quark-gluon language at such low
scales?
What this tells us is that the production of hadrons is driven by 
the strength of the underlying colour fields, the perturbative gluon radiation
probability being a mere tool for quantifying the latter.

There is no need to experimentally verify the MLLA, that is to
check quantum mechanics for ca 200 SF/$Z^0$ (LEP-1). 
However, it is not difficult to imagine {\em a world}\/ without  ``LPHD''.
The QCD string is a nice image for encoding the structure 
of the basic underlying hadronisation pattern (Feynman hot-dog).
What the experimental verification of LPHD tells us, is that hadronisation is 
an amazingly soft phenomenon. 
As far as the global characteristics of final states are concerned,
such as inclusive energy and angular distributions of particle flows,  
there is no visible re-shuffling of particle momenta when  
the transformation from coloured quarks and gluons to 
blanched hadrons occurs.
This means that the QCD string is not a dynamical object, in a sense:
it does not {\em pull}. 
It could, if there were no light quarks around.

\subsection{Some nasty theoretical remarks}
Disproving QCD is no longer in fashion. 
We are now at the stage of trying to understand QCD and learn to 
apply it to a broader range of phenomena. To this end we should ask 
``proper questions'' and use proper tools to avoid confusion. 

%Oscillations of the ratio of the cumulative to factorial moments 
%reported by L3 give an example of an ``improper question'' (IQ). 
%Not that the result was not curious or I had any objections to the
%experimental analysis. The point is that it does not help us much with the
%QCD pursuit: a careful account for energy-momentum conservation seems
%to be enough to explain the phenomenon. 
It is not experimentors' fault that the present-day theory is not
smart enough to extract valuable information from a given
experimental observation. 
Therefore I had better restrain myself from listing ``improper
questions'' (IQs).
% though some of the contributions to HEP-97 gave such an opportunity.
But let me mention one famous case: the long-standing problem of
the ($W\!+\!1$-jet) to ($W\!+\!0$-jet) 
Tevatron ratio may belong to the IQ-club. 
The ratio ($W\!+\!1$-jet)/($W\!+$all) 
would be a safer quantity to check against the fixed-order QCD prediction. 

Two examples of the necessity of ``proper tools''.

There seems to be a problem with electro-production of large-$p_t$ jets.
However, in the region $p_\perp^2\gg Q^2$ 
HERA becomes a ``Tevatron'' with a virtual photon 
substituted for one of the protons. 
Hence, a proper tool here would be merging the parton structure 
of the proton with that of the virtual photon. 
(A general plea: you don't have a proper Monte-Carlo $\neq$ QCD fails!)

Another example is the E706 finding of very broad distributions over the
total transverse momentum of $\pi^0\pi^0$, $\pi^0\gamma$ and $\gamma\gamma$ 
pairs at Fermilab. 
The observed phenomenon is similar to that known for almost 20 years 
in the Drell-Yan pair production.
The proper tool here would be all-order double logarithmic Sudakov
form factor effects as a substitute for the claimed large intrinsic
transverse momentum, $\left\langle k_t\right\rangle>1$~GeV~\cite{E706}.

\section{ICS observables and Confinement}

The much debated problem of power uncertainties in the perturbative expansion 
(the concept of renormalons pioneered by G. 't Hooft and A.~Mueller)
has mutated, all of a sudden, into brave attempts to {\em quantify}\/
power-behaving contributions to various Infrared and Collinear safe
(ICS) observables. This field was initiated and is being pursued by  
Korchemsky and Sterman, Akhoury and Zakharov, Beneke and Braun, 
Nason and Seymour, Shifman, Vainstein and Uraltsev,
Marchesini and Webber, and many others. 

The notion of ICS pQCD predictions ascends to Sterman and Weinberg. 
ICS are the observables that do not contain logarithms of
collinear and/or infrared cutoff $\mu$ in pQCD calculations, and therefore
have a finite $\mu\!\to\!0$ limit. The contribution of small momentum scales to
such quantities should therefore be proportional to $(\mu^2/Q^2)^p$ with
$p>0$, modulo logarithms.  

Simply by examining Feynman diagrams, one can find the powers $p$ for
different observables. This information is already useful: it tells
us how (in)sensitive to confinement physics a given observable is. 
For example, one can compare the performance of different jet 
finding algorithms in this respect to see that hadronisation
corrections to jet rates 
defined with use of the Durham jet-finder are expected to be
smaller ($1/Q^2$)  than those for the JADE algorithms 
($1/Q$)\cite{WParis}.  
More ambitious a programme aims at the {\em magnitudes}\/ of
power-suppressed contributions to hard cross sections/jet observables.

In the last two or three years first steps have been made towards 
a joint technology for triggering and quantifying non-perturbative
effects in ``Euclid-translatable'' cross sections (vacuum condensates)
and, at the same time, in the essentially Minkowskian characteristics 
of hadronic final states.
In the systematic approach known as the ``Wise Dispersive Method''
(WDM)~\cite{WDM} new dimension-full parameters $A_{2p.q}$ emerge that
normalise genuine non-perturbative contributions to dimensionless ICS
observables $V$,
$$
 \delta^{(\mbox{\scriptsize NP})} V \>=\> 
 \sum_{q=0}^{q_m}\rho_q^{(V)} \>
 \left(\ln \frac{Q^2}{\mu^2}\right)^q  \cdot\frac{A_{2p,q}}{(Q^2)^p}+\ldots
$$ 
Perturbative analysis (PT) provides the observable-dependent factors
$\rho_q^{(V)}$ and the leading power $p$. 
The non-perturbative (NP) parameters $A$ are expressed in terms of 
log-moments of the ``effective coupling modification'' 
$$
 A_{2p,q}\>=\> \frac{C_F}{2\pi} \int_0^\infty \frac{dm^2}{m^2} 
 (m^2)^p \left(\ln\frac{m^2}{\mu^2}\right)^{q_m-q} \delta\ae(m^2)\>,
$$
where $p$ is half-integer or $q>0$. 
For different observables, $q_m= 0,1,2$. 
(If $q_m>0$, the contributions combine so as to produce an answer that
does not depend on the arbitrary parameter $\mu$.)

The moments of $\delta\ae(m^2)$ converge at $m^2$ of the order
of $\Lambda^2_{\mbox{\scriptsize QCD}}$. 
The function $\ae$ is related to the standard QCD coupling via the
dispersive integral,
$$
 \alpha_s(k^2) \!=\! \int_0^\infty\!\! \frac{dm^2\>
   k^2}{(m^2+k^2)^2}\>\ae(m^2)\,; \quad
\ae(m^2)\!=\! \frac{\sin(\pi D)}{\pi D}\alpha_s(m^2)\>, \>\> D\equiv
\frac{m^2\,d}{d m^2}\>.
$$
It is thought to be a universal function that characterises,
in an effective way, the strength of the QCD interaction all the way 
down to small momentum scales. Given this universality, 
it becomes possible to {\em predict}\/ the ratios of the $Q^{-2p}$
contributions to observables belonging to the same class $p$.

Those who believe that a school of little fish can be mistaken for
a baby-whale, 
are aware of mounting evidence in
favour of the notion of an infrared-finite QCD coupling\footnote{A certain
irony is necessary since little can be rigourously
proved in the game.}.
Phenomena range from simple estimates of hadron interaction cross sections 
in the Low-Nussinov two-gluon model of the Pomeron, all the way up to
a detailed sophisticated analysis of meson properties in the framework
of the ``relativised'' potential model of Godfrey and Isgur~\cite{MS}. 

Of primary interest are ICS jet-shape observables many of which
exhibit the $1/Q$ leading power corrections, $p=\half$. 
These include the thrust $T$, the so called $C$-parameter, 
invariant jet masses, the jet broadening $B$ ($\ln Q$--enhanced). 
The energy-energy correlation function in $e^+e^-$ annihilation, $\EEC(\chi)$, 
also contains the $1/Q$ confinement contribution (away from the
back-to-back region, $\chi\neq\pi$).

A crucial question is that of marrying PT and NP contributions. 
At the PT level, only the first few orders of the $\alpha_s$ expansion are
available, for most observables up to the next-to-leading $\alpha_s^2$
order. This fact is not too disappointing: a full knowledge of
the PT expansion would be of little help anyway. Indeed, the series
diverge factorially, so that an intrinsic uncertainty of the sum of PT
terms is at the level of that very same power $Q^{-2p}$ (infrared
renormalon ambiguity). 

The price offered in~\cite{DWdiff} for resolving this ambiguity 
was the introduction of a matching infrared scale $\mu_I$, above which
the coupling is well matched by its famous logarithmic PT
expression. The genuine NP $1/Q$ effects can then be expressed in
terms of the effective magnitude of the coupling in the infrared
region, $\bar{\alpha}_0$,  
$$
\bar{\alpha}_0(\mu_I) \equiv \frac{1}{\mu_I}
\int_0^{\mu_I} 
dk\> \alpha_s(k^2)\>.
$$
Experimental analyses carried out by ALEPH, DELPHI, H1, JADE and OPAL
have demonstrated consistency between power terms 
in the {\em mean values}\/ of $1\!-\!T$, $M^2$ and, to a lesser extent, $B$. 
They pointed at the value for
$\bar{\alpha}_0(2\mbox{GeV})$ in the ball-park of
0.5--0.54\footnote{ 
OPAL came up with a smaller value, the reason being an implementation of
a strongly reduced scale of the PT contribution, as substitute for 
part of the power effect. An experimental verification of the
renormalon phenomenon, if you wish.}.

It is clear that the data at {\em smaller}\/ $Q^2$ are more sensitive to
power effects. Hence, a potential advantage of HERA and the necessity of
revisiting JADE and TASSO data. ``Parasitic'' radiative LEP-1 events 
$e^+e^-\to Z^0+\gamma$ also provide a nice opportunity for studying
jet shapes at reduced hardness scales. The L3 collaboration has it
all, but, for the time being, has conservatively restricted itself 
to comparison with the MC models only~\cite{L3rad}.

The most exciting result has emerged from the recent ALEPH study of the thrust 
{\em  distribution}\/ in the energy range from 14 up to 
180~GeV~\cite{ALEPH_610}.  
The quality of the two-parameter fit of the form~\cite{DWdiff}
$$
\frac{d\sigma}{dT}(T) \>=\> 
\left(\frac{d\sigma}{dT}\right)_{\mbox{\scriptsize PT}}
(T-A/Q)\>,
$$
proves to be  {\em better}\/ than 
that of the fits that incorporate MC-generated hadronisation effects! 
The expression in the right-hand-side is the all-order
resummed perturbative spectrum {\em shifted}\/ by  
$A/Q$, that very same NP correction term that appears in 
the mean, $\left\langle1-T\right\rangle$. 
The ALEPH result is
$$
\bar\alpha_0 \>=\>  0.529\pm0.002\pm 0.0034\>, \quad \mbox{with}\>\>
 \alpha_{\MSbar}(M_Z^2)\>=\> 0.1194\pm0.0003\pm0.0035\>. 
$$
A free fit to the power in the form
$$
\left\langle1-T\right\rangle 
=  c_1\alpha_s(Q^2) +c_2\alpha_s^2(Q^2) \>+\> \frac{\mbox{const}}{Q^P}, 
$$
with $c_1$ and $c_2$ the known PT coefficients, yielded 
$P= 0.98 \pm 0.19$, in accord with the theoretical expectation.

It would be premature, however, to celebrate the success of the PT-motivated
approach to the NP-physics outlined above. 
Experimental studies do not yet incorporate the latest theoretical
findings. First is the so-called Milan factor, the two-loop
renormalisation effect that multiplies the
$1/Q$ power terms by the factor 1.8 (for three active 
quark flavours)~\cite{DLMS}. The good news is that this factor is
universal. In spite of this, its inclusion may affect the fits. 

On top of it, in the studies of the {\em jet broadening}\/ (H1, JADE) 
the wrong relation $\rho^{(B)}_1=\rho^{(1-T)}_0$, instead of 
the correct  $\rho^{(B)}_1=\half\rho^{(1-T)}_0$, was being used 
(stemming from an unfortunate misprint in the theoretical paper).
Moreover, an improved PT description of the $B$-distribution 
is now available~\cite{DLMSbroad} and should be implemented. 

These reservations do not undermine the main amazing finding that a
pure perturbative analysis is capable of predicting the power of the
$Q$-dependence and the magnitude of genuine confinement effects 
in hard observables in general, and in jet shapes in particular. 

With the notion of an infrared-finite coupling 
gaining grounds, we shall be able to speculate about the 
characteristic QCD parameter, 
$$
 \frac{\alpha_s}{\pi} \>\simeq\> 0.16\>, \qquad
\mbox{versus} \quad  \frac{\alpha_{\mbox{\scriptsize crit}}}{\pi} 
= \frac1{C_F}\left(1-\sqrt{\frac{2}{3}}\,\right)\approx 0.137\>,
$$
being sufficiently small to allow the application 
of perturbative language, at least semi-quantitatively, 
down to small momentum scales.
At the same time, it appears to be sufficiently large to activate the Gribov
super-critical light-quark confinement mechanism~\cite{BH}.

\section{Conclusions}
\subsection{Theory}

QCD is an infrared-unstable theory.
Physical states are Swedish miles\footnote{1 Swedish mile = 10 km}
away from the fundamental objects making up the QCD Lagrangian. 
In such circumstances we had better be sceptical and put under Cartesian
scrutiny our field-theoretical concepts and tools. 
In particular, we expect quark and gluon Green's functions 
to have weird analytic properties 
as they ought to describe {\em decaying}\/ objects, in a rather
unprecedented way. 
This fact makes the concept of ``Euclidean rotation'' far from secure
and, in principle, undermines the familiar statistical
mechanics substitute for Minkowskian field theory (read: lattice).   

To understand the structure of the QCD vacuum and that of hadrons in
the real world, we have to address the general problem of binding 
massless particles. 
The Gribov super-critical confinement remains, at present, the only 
dynamical mechanism proposed for that. 

The 13 puzzles will stay with us for a while longer.

\subsection{Phenomenology}

The quantitative theory of hadrons, which theoreticians ought to be looking
for, gets more and more restricted by the findings of our
experimenting colleagues. 
The news is, that the small-distance (pQCD) approach, 
using quarks and gluons, works too often too well.
Hadronisation effects, when viewed {\em globally}\/ seem to behave 
surprisingly amicably: they either stay {\em invisible}\/ 
(inclusive energy and angular hadron spectra) 
or can be quantified (power effects). 

A pQCD-motivated technology for triggering and quantifying genuine 
non-per\-tur\-bative (confinement) effects is under construction.
These effects show up as power-behaving contributions to
Infrared/Collinear-Safe observables, and jet shapes in particular.
From within perturbation theory the leading powers can be detected,
and the {\em relative}\/ magnitude of power terms predicted.
The absolute values of new dimensional parameters, 
which we find phenomenologically these days, can  be related to the
shape of the effective interaction strength (effective QCD coupling)
in the infrared region, 
$\left\langle \alpha_s/\pi\right\rangle\sim 0.2$.

\subsection{Experiment (``what am I doing this for?'')}

The epoch of basic QCD tests is over.
Today's quest is to understand hadron structure via hadron interactions.  
The major goal is to study the interface between small and large distances.

The best laboratory for that is ``almost-photo-production'' at
HERA, that is the interaction of small-virtuality photons, 
$0<Q^2<4\,\mbox{GeV}^2$, or so, with protons and, hopefully, nuclei. 

Diffraction phenomena also target smaller-than-typical
hadronic states (e.g., $t$- and $Q^2$--dependence of vector meson 
photo/electro-production). 

Studying Drell-Yan pairs, $J/\psi$ and $\Upsilon$ in $pp$ ($p\bar{p}$),
$pA$ and $AB$ interactions remains a top priority. In addition to the
total production cross sections and the $p_t$--distributions, various 
correlation experiments are extremely informative (a famous
example being the NA~50 study of the $J/\psi$ yield as a function of
accompanying hadronic activity, $E_T$, in ion-ion collisions).

Jet studies looking for differences in hadron abundances between quark-
and gluon-initiated jets should be pursued. Similarity of the yield of
hadrons of different species in $q$ and $g$ jets in the ``sea''
region, if confirmed, would be of major importance for understanding
hadronisation dynamics. 

Differential jet rates and internal jet-substructure of jets in hard
interactions also provide a handle on the soft-hard interface, when
one increases jet resolution (by decreasing $y_{\mbox{\scriptsize  cut}}$). 
To look for genuine confinement effects in such observables, special
care should be taken to preserve, as much as possible, the
correspondence between parton and hadron ensembles at the perturbative
level. To this end the recently proposed modified Durham jet finder,
the so-called Cambridge jet algorithm, will have to be used.

\subsection{Overall}
QCD is on the move, and the pace is good.

\vspace{2 cm}

\noindent {\bf Acknowledgements}

\noindent
I am grateful to Marek Karliner for the opportunity to visit  
Tel Aviv University where this talk was prepared. 
I enjoyed the hospitality of and invaluable on-line help from my dear friends 
Halina Abramowicz and Aharon Levy. 
I would also like to thank Lenya Frankfurt for teaching me high energy
nuclear physics.

\vspace {2 cm}
%%\newpage
\def\labelenumi{[\arabic{enumi}]}

\noindent
{\Large\bf References}
\begin{enumerate}
\item\label{Torah} ``nothing to add, nothing to subtract''; 
                     Torah (about Torah).
\item\label{Mueller} A.H. Mueller, Phys.Lett. { B396} (1997) 251.
% hep-ph/9612251.
\item\label{CCnpt} G. Camici and M. Ciafaloni, 
                   Phys.Lett. { B395} (1997) 118.
\item\label{CCint} 
%% G. Camici and M. Ciafaloni,  
%%                   Phys.Lett. 
ibid. { B412} (1997) 396.
\item\label{BV} J. Bl\"umlein and A. Vogt, Phys.Rev. D57 (1998) 1.
\item\label{Marcello_pc}  M. Ciafaloni, private communication.
\item\label{ZEUS_MN} ZEUS contribution 659.
\item\label{Delduca} V. Del Duca and C.R. Schmidt,
  Nucl.Phys.Proc.Suppl. { 39BC} (1995) 137. 
\item\label{D0azim} D0 contribution 087.
\item\label{FP} E.L. Feinberg and I.Y. Pomeranchuk,
  Suppl.Nuovov.Cim. { III} (1956) 652.
\item\label{GW} M.L. Good and W.D. Walker, Phys.Rev. { 120}
(1960) 1857.
\item\label{MP} H. Miettinen and J. Pumplin, 
%% Phys.Rev. {\bf D18} (1978) 1696; 
Phys.Rev.Lett. { 42} (1979) 204.
\item\label{FRS} L. Frankfurt, A. Radyushkin and M. Strikman,
Phys.Rev. { D55} (1997) 98, and references therein. 
\item\label{DELPHI_545} DELPHI contribution 545.
\item\label{OPALglue} OPAL contribution 181.
\item\label{H1_251} H1 contribution 251
\item\label{zeus_662} ZEUS contribution 662.
\item\label{L3lam} L3 contribution 506.
\item\label{OPALlam} OPAL contribution 192.
\item\label{CDF} CDF contribution 562.
\item\label{KO} review: V.A. Khoze and W. Ochs, 
 Int.J.Mod.Phys. { A12} (1997) 2949.
\item\label{E706} E706 contribution 256.
\item\label{WParis} B.R. Webber, talk at the Workshop on Deep
  Inelastic Scattering and QCD, Paris, April 1995; hep-ph/9510283.
\item\label{WDM}  Yu.L.\ Dokshitzer, G. Marchesini and  B.R.\ Webber, 
 Nucl.Phys. { B469} (1996) 93.
\item\label{MS} A.C. Mattingley and P.M.  Stevenson, 
 Phys. Rev. { D49} (1994) 437,
%% ; hep-ph/9307266, 
and references therein. 
\item\label{DWdiff} Yu.L. Dokshitzer and B.R. Webber, Phys.Lett. 
   { B352} (1995) 451; ibid.  { B404} (1997) 321.
%% POWERS
%\item\label{H1_253}     H1 contribution 253 
\item\label{L3rad} L3 contribution 499.
\item\label{ALEPH_610}  ALEPH contribution 610.
%\item\label{DELPHI_544} DELPHI contribution 544  %% also MLLA spectrum
\item\label{DLMS} Yu.L. \ Dokshitzer, A.\ Lucenti,  
G.~Marchesini and G.P.~Salam,  hep-ph/9707532. 
\item\label{DLMSbroad} ibid.  hep-ph/9801324.
\item\label{BH} V.N. Gribov, Physica Scripta {T15} (1987) 164;
                Lund preprint LU 91--7 (1991).
\end{enumerate}

\end{document}